\documentclass[a4paper,11pt,fleqn]{article}

\usepackage[ansinew]{inputenc}
\usepackage{hyperref}
\usepackage[mathscr]{eucal}
\usepackage{graphicx}
\usepackage[font=footnotesize,labelsep=period,justification=justified,figureposition=bottom,tableposition=top]{caption}
\usepackage{subcaption}
\usepackage{multirow}
\usepackage{amsmath,amssymb,amsthm}
\usepackage{comment}
\allowdisplaybreaks

\hypersetup{colorlinks, linkcolor=blue, citecolor=blue, urlcolor=blue}
\newcommand{\todo}[1][\null]{\ensuremath{\clubsuit}}

\newcommand{\noprint}[1]{}

{\theoremstyle{definition}

\newtheorem*{remark*}{Remark}
}

\newcommand{\checked}[1][\null]{\ensuremath{\boldsymbol{\surd}}}

\begin{document}

\par\noindent {\LARGE\bf
M-ENIAC: A machine learning recreation of the first successful numerical weather forecasts
\par}

\vspace{4mm}\par\noindent {\large
R\"udiger Brecht$^{\dag}$ and Alex Bihlo$^\ddag$
\par}

\vspace{4mm}\par\noindent{\it
$^{\dag}$Department of Mathematics, University of Hamburg, Hamburg, Germany
}

\vspace{2mm}\par\noindent{\it
$^{\ddag}$ Department of Mathematics and Statistics, Memorial University of Newfoundland,\\
$\phantom{^{\ddag}}$~St.\ John's (NL) A1C 5S7, Canada
}

\vspace{2mm}\par\noindent {\it
\textup{E-mail:} rbrecht@uni-hamburg.de, abihlo@mun.ca
}\par

\vspace{12mm}\par\noindent\hspace*{10mm}\parbox{140mm}{\small
In 1950 the first successful numerical weather forecast was obtained by solving the barotropic vorticity equation using the Electronic Numerical Integrator and Computer (ENIAC), which marked the beginning of the age of numerical weather prediction. Here, we ask the question of how these numerical forecasts would have turned out, if machine learning based solvers had been used instead of standard numerical discretizations. Specifically, we recreate these numerical forecasts using physics-informed neural networks. We show that physics-informed neural networks provide an easier and more accurate methodology for solving meteorological equations on the sphere, as compared to the ENIAC solver.
\par}\vspace{7mm}

\section{Introduction}\label{sec:IntroductionBVE}

Numerical weather prediction is the backbone of modern meteorology~\cite{baue15a}. Without numerical weather forecasts it would be impossible to predict the weather in advance with as much accuracy as is possible today. While numerical weather prediction is being taken for granted nowadays, it was first shown feasible by the seminal work of Charney, Fj{\o}rtoft and von Neumann~\cite{char50a}, although the \textit{dream} of numerical weather prediction is much older, dating back to Richardson~\cite{rich22a} more than 100 years ago. The first successful numerical weather forecast was carried out on the \textit{Electronic Numerical Integrator and Computer} (ENIAC), using a highly simplified version of the governing equations of the atmosphere. This simplified equation, the so-called barotropic vorticity equation, was solved using a straightforward finite-difference method, and the resulting accuracy of the forecasts was indeed quite underwhelming~\cite{lync08a}. Still, in comparison to the numerical forecast attempted by Richardson thirty years earlier, the forecasts by Charney, Fj{\o}rtoft and von Neumann were a resounding success, that eventually lead to the quiet revolution of numerical weather prediction~\cite{baue15a}.

The ENIAC forecasts were recreated in the paper~\cite{lync08a}, and later carried out on a cellphone~\cite{lync08b} to showcase the dramatic improvements of processing powers, and the potential that this holds for new avenues for numerical weather forecasting. With the recent breakthrough in training deep neural networks~\cite{kriz12a}, and the enormous interest in both using neural networks for solving differential equations~\cite{rais18a} and meteorology in general, we argue it is natural to revisit the ENIAC integrations again, this time using machine learning.

More specifically, in this paper we ask the following speculative question: \textit{How would the first successful numerical weather forecasts have turned out, had machine-learning based numerical solvers be used instead of classical numerical integration?} We refer to this task as M-ENIAC, a machine learning recreation of the ENIAC forecasts. 

While the recreation of the ENIAC forecasts using neural networks may seem contrived, we do believe this is a timely problem to consider for several reasons. As most other fields of the mathematical sciences, meteorology has also seen a substantial increase of interest in machine learning and deep learning. While neural networks have been considered in this field as early as the 1990s, see~\cite{gard98a} for a historical review, the most recent surge in breakthroughs using deep neural networks has sparked a renewed interest in these networks in meteorology in the past few years, with several groups working on integrating deep neural networks into operational weather forecasting. Indeed, recently it was shown that a Transformer based weather forecasting model can outperform the state-of-the art ECMWF Integrated Forecasting System~\cite{bi22a}.

In parallel to the breakthroughs of neural networks in computer vision, they have also been proposed as an alternative to traditional numerical schemes for solving differential equations~\cite{laga98a,rais18a}. While this method has seen an enormous surge in interest in just a few years, as of now this field is still in its infancy. Specifically, there are almost no applications of this method in meteorology, having only been applied to solve some of the standard benchmarks for the shallow-water equations on the sphere~\cite{bihl22a}. Importantly, to the best of our knowledge, this method has not yet been used for weather forecasting using real weather data. As such, given the rise in interest in neural networks both for weather forecasting, and for solving differential equations, we believe it is fitting to return to the problem of weather forecasting with the barotropic vorticity equation. We will highlight some parallels between the earliest successful weather forecast using finite differencing, and the state-of-the-art in using neural networks for solving the same equation, which include both performance and accuracy issues, but also hint at some potential benefits of neural network based numerical discretization methods over more traditional approaches.

The further organization of this paper is as follows. In the following Section~\ref{sec:PreviousWorkBVE} we review the relevant literature on meteorological applications of machine learning and how neural networks can be used for solving differential equations. In Section~\ref{sec:BVEBVE} we introduce the barotropic vorticity equation and discuss its underlying assumptions along with its limitations for weather forecasting. This section also contains the description of how neural networks can be used to solve this equation. The main Section~\ref{sec:PINNBVE} presents the numerical results of integrating the vorticity equation on the sphere using physics-informed neural networks. We wrap up the paper with a summary of our results presented in Section~\ref{sec:ConclusionsBVE}.

\section{Previous work}\label{sec:PreviousWorkBVE}

Numerical weather prediction was first proposed by Richardson in his famous 1922 book~\cite{rich22a}. While impractical at the time, Richardson's ideas on numerically solving the governing equations of atmosphere--ocean dynamics would eventually be put into practice several decades later. The first actual successful (but highly simplified) numerical weather forecast was carried out by Charney, Fj{\o}rtoft and von Neumann and described in the landmark paper~\cite{char50a}. Here, the authors numerically solved the barotropic vorticity equation on a polar stereographic plane over North America and the North Atlantic. While being worse than the persistence prediction for three out of four 24 hour forecasts carried out~\cite{lync08a}, this work sparked considerable investments into the field that over the decades to follow became a resounding success story~\cite{baue15a}. 

Today numerical weather prediction is carried out largely using traditional numerical methods, in particular finite difference, finite element or pseudospectral discretization methods~\cite{durr10a}. With the emergence of modern deep learning, there has been considerable interest in applying machine learning to meteorology and climate science in recent years. This includes all aspects of these fields, such as data assimilation~\cite{geer21a}, weather forecasting~\cite{bi22a,bihl21a,path22a,weyn19a}, ensemble forecasting~\cite{bihl21a,brech23a, sche18a}, weather nowcasting~\cite{xing15a}, downscaling~\cite{moua17a}, and subgrid-scale parameterization~\cite{gent18a}. 

While all of these references focus on data-driven approaches, there have also been first examples for solving the governing equations directly using neural networks~\cite{bihl22a}. This latter approach, dubbed \textit{physics-informed neural networks} in the paper~\cite{rais18a}, has seen considerable interest in the mathematical sciences lately. Physics-informed neural networks aim at representing the solution of a differential equation using a neural network that is trained to satisfy the underlying equation at a collection of collocation points. This method hence bears some resemblance with Hermite interpolation, although it does not aim to fit derivative data only but an entire differential equation. For computing derivatives, the method relies on automatic differentiation~\cite{bayd18a} rather than numerical differencing, which is directly available in modern deep learning frameworks such as \texttt{TensorFlow}, \texttt{PyTorch} or \texttt{JAX}. This allows computing derivatives in single points, without having to rely on defining a computational mesh first. The method thus is entirely meshless, which makes it potentially attractive in meteorology, which had to devise a plethora of grids~\cite{stan12a} for solving the governing equations of hydro-thermodynamics on a sphere using standard mesh-based discretization methods.

While the main idea behind physics-informed neural networks is quite elegant, in practice there are several issues that will have to be overcome to make them applicable for the atmospheric sciences. Physics-informed neural networks tend to converge to the wrong solution for longer training times~\cite{wang23a,wang22a}, and they typically, as of now, cannot obtain the same level of error as traditional numerical methods can achieve, see for example~\cite{penw23a} for a study on various variations of physics-informed neural networks and their associated error levels. Succinctly speaking, it is  often challenging to show that physics-informed neural networks converge at all. This is likely not a theoretical shortcoming, but a practical one, as the optimization methods used for minimizing the physics-informed loss function may simply not yield an accurate enough minimum, resulting in rather large numerical errors of the associated numerical solution. Using tailored optimization methods can vastly improve this situation~\cite{bihl23a}. Physics-informed neural networks also take much longer to train in practice than solving the same equations using a traditional numerical method. For example, the physics-informed neural networks being trained here took about four hours each on a single NVIDIA RTX8000 GPU, while the re-created ENIAC solver from~\cite{lync08a} solves the same problem in a fraction of a minute. For meteorological applications it should also be stressed that as of now we are not aware of any example of a physics-informed neural network that could numerically preserve conservation laws up to machine precision, although some progress is being made in this direction as well~\cite{jagt20a}.

All these issues bear some remarkable resemblance to the early days of numerical weather forecasting. The numerical scheme used by~\cite{char50a} was neither particularly accurate~\cite{lync08a}, nor was it conservative, with the first conservative numerical formulation for the vorticity equation only being proposed about fifteen years later~\cite{arak66a}. The actual integrations by~\cite{char50a} also took some 24 hours to carry out for a 24 hour prediction, meaning this forecast was able to just keep in pace with the evolution of the atmosphere itself; it would not have been useful for operational weather forecasting. The same would likely hold true as of now for physics-informed neural networks for more complicated models from atmosphere--ocean dynamics, although parallelization of the training of these networks on multi-GPU machines would be possible. 

\section{The barotropic vorticity equation on the sphere}\label{sec:BVEBVE}

Here we describe the form of the vorticity equation on the sphere which we will solve using physics-informed neural networks.

\subsection{Model formulation}

The barotropic vorticity equation is a single prognostic equation for the geopotential height of the atmosphere. It is derived from the two-dimensional Euler equations under the assumption of non-divergence of the velocity field, which allows recasting of these equations using the stream function. On the rotating sphere, the barotropic vorticity equation is given by
\begin{align}\label{eq:VorticityEquation}
\begin{split}
\zeta_t + \frac{1}{a^2\cos\varphi}(\psi_\lambda\zeta_\varphi - \psi_\varphi\zeta_\lambda) + \frac{2\Omega}{a^2}\psi_\lambda = 0,\quad
\zeta = \frac{1}{a^2\cos\varphi}\left(\frac{1}{\cos\varphi}\psi_{\lambda\lambda} + (\cos\varphi \psi_\varphi)_\varphi\right),
\end{split}
\end{align} 
where $\lambda\in[-\pi,\pi]$ and $\varphi\in[-\pi/2,\pi/2]$ are the longitude and latitude, respectively, and subscripts denote derivatives with respect to the associated variable. Here, $\psi=\psi(t,\lambda,\varphi)$ is the streamfunction generating non-divergent flow, $\zeta$ is the associated vorticity, related to the streamfunction by the Laplacian in spherical coordinates, $\Omega$ is the angular velocity and $a$ the radius of Earth. The stream function is related to the geopotential height $z$ via $\psi=gz/f_0$, with $g$ being the gravitational acceleration and $f_0=2\Omega\sin\varphi_0$ being the Coriolis parameter at a constant reference latitude. For more details on this model, consult the textbooks~\cite{holt04a,sato04a}.

The barotropic vorticity equation is an exceptional model in dynamic meteorology as it condenses down the evolution of the atmosphere into a single prognostic equation. Derived in~\cite{charn48a} as an approximation of the primitive equations, the barotropic model eliminates several complications of these equations, such as the presence of both fast and slow moving waves, by filtering out the fast gravity waves and only retaining the slow Rossby waves. The batrotropic vorticity equation is also easier to integrate as only a single variable (the geopotential height or pressure field) is needed, rather than both pressure and wind fields. This was the model chosen for the ENIAC integrations~\cite{char50a}, see~\cite{lyne02a} for an excellent historical review and recreation of these forecasts. What was forecast was the height $z$ of the 500 hPa pressure surface, the so-called \textit{geopotential (height field)}, instead of the stream function, and the goal was to compare the forecast of $z$ after 24 hours to the observed geopotential height.

We note here that the original ENIAC integrations~\cite{char50a} were done in Cartesian coordinates, using a polar stereographic projection of the spherical geopotential height field onto a two-dimensional plane. We speculate that this was done largely out of necessity, since solving~\eqref{eq:VorticityEquation} requires time-stepping of the vorticity and a subsequent numerical solution of the Poisson equation for the stream function, which is a non-trivial task on the sphere. On the plane this Poisson equation can be solved using Fourier series, on the sphere this solution would require the use of spherical harmonics. Since the original input data is of course given on the sphere as the geopotential height $z$, and as we shall review below, no issue with solving Eq.~\eqref{eq:VorticityEquation} directly using neural networks arise, we will solve this equation as given, and will only project the results onto the used two-dimensional plane for comparison purposes against the original ENIAC integrations.

\subsection{Data}

The initial conditions and verifying analysis for the geopotential height were obtained from the NCEP--NCAR reanalysis data~\cite{kaln96a}. This reanalysis data is available from January 1948 onward and thus covers the four ENIAC test cases, which were (i) January 5th, 1949, (ii) January 30th, 1949, (iii) January 31st, 1949, and (iv) February 13th, 1949. The spatial resolution of the data is $2.5^\circ\times 2.5^\circ$, equating to $144\times73$ grid points. We obtained the geopotential height $z$ of the 500 hPa pressure surface. As indicated above, the stream function itself is related to the given geopotential height by $\psi=g z/f_0$. We set the reference latitude in the Coriolis parameter $f_0$ to $\varphi_0=\pi/4$, or 45 degree North~\cite{holt04a}.

\subsection{Solving the vorticity equation using neural networks}

In the following, we denote by $\Delta(t,\lambda,\varphi,\psi_{(3)}(t, \lambda, \varphi)) = 0$ the equation obtained by substituting the vorticity $\zeta$ in Eq.~\eqref{eq:VorticityEquation} into the first equation and converting it into a third-order partial differential equation for the streamfunction $\psi$. Here, we denote with $\psi_{(3)}$ all partial derivatives of the streamfunction with respect to the independent variables of order not greater than three. The ENIAC integrations solved the vorticity equation for 24 hours, or $t_f=86400$ seconds.

For solving the barotropic vorticity equation $\Delta = 0$ using physics-informed neural networks we parameterize the solution as a neural network of the form $\psi^{\boldsymbol{\theta}}=\mathcal N^{\boldsymbol{\theta}}(t,\lambda, \varphi)$, where $\boldsymbol{\theta}$ are the weights of the neural network. That is, the input to the neural network is a point $(t,\lambda, \varphi)$ in three-dimensional space-time, and the output is the scalar $\psi^{\boldsymbol{\theta}}$, which should be an approximation to the true streamfunction at the given point. 

The weights of this neural network are then found by minimizing the physics-informed loss function
\begin{subequations}
\begin{equation}\label{eq:PINNLoss}
\mathcal{L}(\boldsymbol{\theta}) = \gamma_{\Delta}\mathcal{L}_\Delta(\boldsymbol{\theta}) + \gamma_{\rm i}\mathcal L_{\rm i}(\boldsymbol{\theta}),
\end{equation}
with the equation loss and initial value loss being defined as the mean-squared errors
\begin{align}
\begin{split}
\mathcal L_\Delta(\boldsymbol{\theta}) &= \frac{1}{N_\Delta}\sum_{j=1}^{N_{\Delta}}|\Delta(t^j_\Delta, \lambda^j_\Delta, \varphi^j_\Delta, \psi_{(3)}^{\boldsymbol{\theta}}(t^j_\Delta, \lambda^j_\Delta, \varphi^j_\Delta))|^2,\\
\mathcal L_{\rm i}(\boldsymbol{\theta}) &= \frac{1}{N_{\rm i}}\sum_{j=1}^{N_{\rm i}}|\psi^{\boldsymbol{\theta}}(0,\lambda^j_{\rm i}, \varphi^j_{\rm i})-\psi_0(\lambda^j_{\rm i}, \varphi^j_{\rm i})|^2,
\end{split}
\end{align}
\end{subequations}
and with $\gamma_{\Delta},\gamma_{\rm i}\geqslant0$ being positive scaling constants. These losses are minimized over the set of collocation points $\{t^j_\Delta, \lambda^j_\Delta, \varphi^j_\Delta\}_{j=1}^{N_\Delta}$ covering the spatio-temporal domain over which the vorticity equation should be solved, and $\{0, \lambda^j_{\rm i}, \varphi^j_{\rm i}\}_{j=1}^{N_{\rm i}}$ covering the spatial domain at the initial time $t=0$. Here, $\psi_0=\psi(0,\lambda, \varphi)$ is the initial condition for which the vorticity equation is being solved. 

The physics-informed loss function adjusts the weights $\boldsymbol{\theta}$ of the neural network in such a manner that both the differential equation and the associated initial condition hold, upon which point $\psi^{\boldsymbol{\theta}}=\mathcal N^{\boldsymbol{\theta}}(t,\lambda, \varphi)$ will provide a numerical approximation to the true solution $\psi=\psi(t,\lambda,\varphi)$. Here we use a simple feedforward neural network for $N^{\boldsymbol{\theta}}$. Specifically, all neural networks trained in the following have a total of 10 hidden layers with 80 units per layer, using the hyperbolic tangent activation function. The collocation points $\{t^j_\Delta, \lambda^j_\Delta, \varphi^j_\Delta\}_{j=1}^{N_\Delta}$ for the vorticity equation were sampled from the temporal-spatial domain $\Omega=[0,86400]\times[-\pi,\pi]\times[-\pi/2,\pi/2]$ such that the spatial points covering the sphere were uniformly distributed, i.e.\ fewer points going either north or south from the equator towards the poles. This was realized by sampling an $(N_\Delta, 3)$-array of points from the interval $[0,1]$ using Latin hypercube sampling~\cite{stei87a}. Each resulting sampled point $(t^*,\lambda^*,\varphi^*)$ was then mapped to the domain $\Omega$ using the point transformation
\[
t_\Delta = t^*,\quad \lambda_\Delta = -\pi + 2\pi\lambda^*,\quad \varphi_\Delta = -\frac{\pi}{2} + \arccos(1-2\varphi^*),
\]
see~\cite{simo15a} for further details. For generating the initial value collocations points we proceed in the same manner, except for only sampling  $(N_{\rm i}, 2)$-tuples of points, with each sampled point being $(\lambda^*,\varphi^*)$ and $t^*=0$ for all such pairs of spatial points. We use $N_\Delta=100,000$ and $N_{\rm i}=10,000$ collocation points for the differential equation and initial value loss, respectively. These values were chosen to roughly equate the spatial resolution of the gridded reanalysis data, $144\times73=10,512$, and factoring in the time step of 2 to 3 hours chosen in~\cite{char50a}, which would yield a total of 84,096 to 126,144 computational points over the global spatio-temporal domain. While a direct comparison between the classical numerical method used in~\cite{char50a} and neural network based methods is of course challenging, the chosen values at least make the used computational grids comparable. We use bilinear interpolation from the given initial reanalysis geopotential height on the latitude--longitude grid to sample the initial condition at the required collocation points.  

Neural networks are known to be sensitive if their input range is not normalized. Thus, we normalize the time values from $[0,86400]$ to $[-1,1]$ in a custom normalization layer to the network. Since the hidden layers of our network produce values between $-1$ and $1$ and the values of the unscaled stream function are large, of the order of $10^8$, we pass the output of the network to a custom scaling layer, that multiplies the predicted value with $\Psi=a^2/t_{\rm f}$. To non-dimensionalize the loss function, we set $\gamma_{\Delta}=(a^2t_{\rm f}/\Psi)^2$ and $\gamma_{\rm i} = 10/\Psi^2$. This allows us to work with the equation in the form~\eqref{eq:VorticityEquation}, without having to non-dimensionalize the equation itself.

We use the same approach as in~\cite{bihl22a} to enforce the boundary conditions on the sphere, namely passing the input $(\lambda, \varphi)$-coordinates into a custom coordinate-transform layer that computes $(\cos\varphi\cos\lambda,\cos\varphi\sin\lambda,\sin\varphi)$, meaning it embeds the sphere into $\mathbb{R}^3$ via an implicit use of Cartesian coordinates $\mathbf{r}=(x,y,z)$, so that the neural network solution is represented as $\psi^{\boldsymbol{\theta}}=\psi^{\boldsymbol{\theta}}(t,\mathbf{r}(\lambda,\varphi))$. This is done to simplify the loss function~\eqref{eq:PINNLoss} which otherwise would also require an explicit boundary value loss. 

\subsection{Inplementation}

The method described in the previous subsection was implemented using \texttt{TensorFlow} 2.12, with training being done on a single NVIDIA RTX8000 GPU. Training for each model took about 4 hours. The loss function was minimized using a standard Adam optimizer~\cite{king14a} with a learning rate of $10^{-3}$, using a mini-batch size of 1000 collocation points for the differential equation loss and 100 collocation points for the initial value loss per batch. Training was done using single precision arithmetic to reduce the overall computational cost. The codes will be made available on GitHub upon publication of this paper\footnote{\url{https://github.com/abihlo/MENIAC}}.

\section{Numerical results}\label{sec:PINNBVE}

We present here the numerical results of our machine learning recreation of the ENIAC forecasts. The evaluation is done on the original ENIAC domain, which encompasses North America and the North Atlantic.
\subsection{Error measures}
As error measures we choose the same measures as reported in~\cite{lync08a}, which are the
\begin{align*}
    &\mathrm{Mean\ error} = \frac{1}{n}\sum_{i=1}^n(\psi^{\rm fcst}_i - \psi^{\rm gt}_i)\\ 
    &\mathrm{RMS\ error} = \sqrt{\frac{1}{n}\sum_{i=1}^n(\psi^{\rm fcst}_i - \psi^{\rm gt}_i)^2}\\
    &\mathrm{S1\ score} = \frac{\sum_{\rm adjacent\ pairs}|\Delta \psi^{\rm fcst} -\Delta \psi^{\rm gt}|}{\sum_{\rm adjacent\ pairs} \max(|\Delta \psi^{\rm fcst}|, |\Delta \psi^{\rm gt}|)}\times 100
\end{align*}
where $\psi^{\rm fcst}$ is the forecast stream function and $\psi^{\rm gt}$ is the true stream function after 24 hours, and $\Delta \psi^{\rm fcst}$ and $\Delta \psi^{\rm gt}$ denote the spatial differences between adjacent grid points of the forecast and ground truth stream functions, respectively. 

The S1 score is a measure of the accuracy of the gradients of forecasts, which are important for meteorological parameters such as geopotential or pressure, as their gradients are directly related to the wind fields~\cite{holt04a}. Perfect forecasts will have an S1 score of zero. For more details on these verification measures, see~\cite{wilk11a}.
\subsection{Evaluation}
We choose two baseline forecast models here, the original ENIAC integrations~\cite{char50a}, obtained from the recreation of these forecasts in~\cite{lync08a}, and the persistence prediction. The persistence prediction predicts that the final stream function field is exactly the same as the initial field, i.e.\ that there is no temporal evolution over the forecast horizon. Note that the values for the ENIAC and persistence predictions are the same as in Table 1 of~\cite{lync08a}.

\begin{table}[!ht]
\centering
\caption{Verification measures for the three forecast models, physics-informed neural networks (MENIAC), the original ENIAC forecasts (ENIAC) and the persistence forecast (Pers). Reported are mean error, root mean squared error and S1-score. Bold indicates the best values.}
\label{tab:VerificationEniac}
\begin{tabular}{c||ccc||ccc||ccc}
     & \multicolumn{3}{c||}{Mean error}                                                 & \multicolumn{3}{c||}{RMS}                                          & \multicolumn{3}{c}{S1 score}                                              \\ \hline
Case & \multicolumn{1}{c|}{MENIAC}          & \multicolumn{1}{c|}{ENIAC}  & Pers         & \multicolumn{1}{c|}{MENIAC}          & \multicolumn{1}{c|}{ENIAC}  & Pers & \multicolumn{1}{c|}{MENIAC}          & \multicolumn{1}{c|}{ENIAC}          & Pers \\ \hline
1    & \multicolumn{1}{c|}{-15}         & \multicolumn{1}{c|}{56}  & \textbf{-9} & \multicolumn{1}{c|}{\textbf{66}} & \multicolumn{1}{c|}{113} & 95  & \multicolumn{1}{c|}{\textbf{47}} & \multicolumn{1}{c|}{61}          & 62  \\
2    & \multicolumn{1}{c|}{-8}          & \multicolumn{1}{c|}{31}  & \textbf{6}  & \multicolumn{1}{c|}{\textbf{85}} & \multicolumn{1}{c|}{99}  & 115 & \multicolumn{1}{c|}{48}          & \multicolumn{1}{c|}{\textbf{46}} & 63  \\
3    & \multicolumn{1}{c|}{\textbf{12}} & \multicolumn{1}{c|}{-35} & 20          & \multicolumn{1}{c|}{\textbf{78}} & \multicolumn{1}{c|}{93}  & 89  & \multicolumn{1}{c|}{\textbf{44}} & \multicolumn{1}{c|}{46}          & 58  \\
4    & \multicolumn{1}{c|}{-4}          & \multicolumn{1}{c|}{39}  & \textbf{1}  & \multicolumn{1}{c|}{\textbf{68}} & \multicolumn{1}{c|}{82}  & 81  & \multicolumn{1}{c|}{41}          & \multicolumn{1}{c|}{\textbf{39}} & 50 
\end{tabular}
\end{table}

Table~\ref{tab:VerificationEniac} shows that the recreated ENIAC integrations using physics-informed neural networks outperform the original ENIAC integrations in all four cases using the RMS error measure, and importantly also outperform persistence in all four cases. Note that this was not the case for the original ENIAC integrations, for which only one forecast was better than persistence in terms of the RMS error. The physics-informed neural network predictions also outperform persistence in one case using the mean error measure, while the original ENIAC integrations did not outperform persistence in a single case in this error measure. We also note that physics-informed neural networks outperform the ENIAC integrations in all cases in the mean error measure. The S1 score indicates that both ENIAC and MENIAC outperform persistence, with MENIAC outperforming ENIAC in two of the four cases, and coming close to matching the S1 scores for the remaining two cases.

\begin{figure}[!ht]
    \centering
    \includegraphics[width=\textwidth]{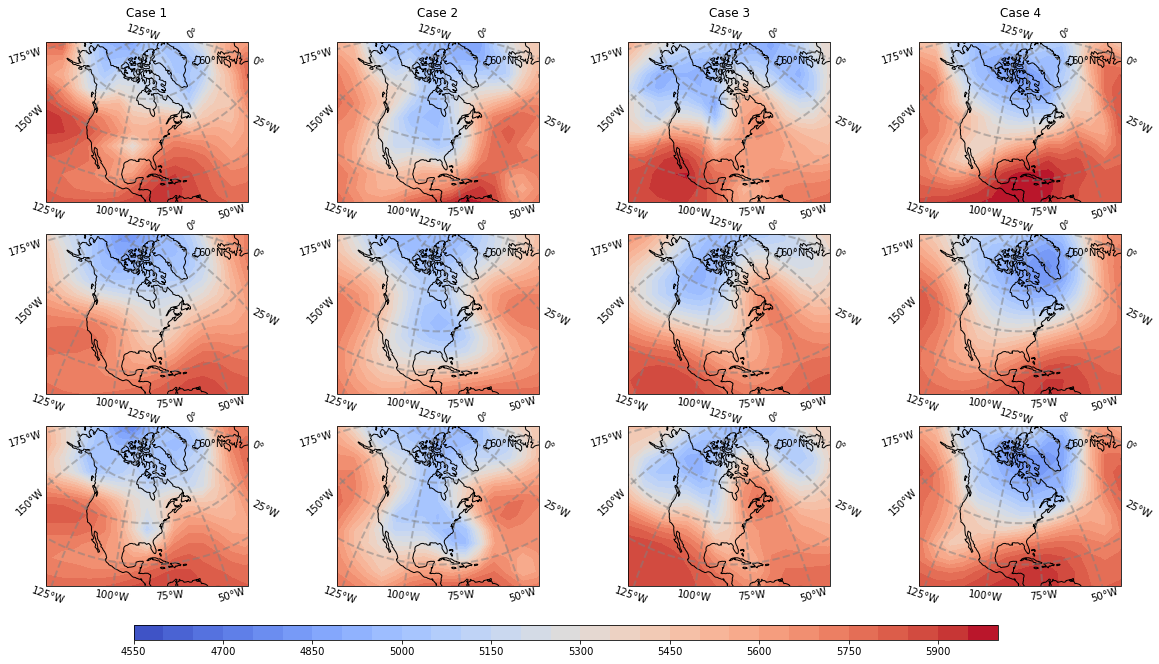}
    \caption{Forecast of the 500 hPa geopotential height. \textit{Top row:} Original ENIAC integrations, \textit{Middle row:} M-ENIAC, \textit{Bottom row:} Ground truth.}
    \label{fig:compare_z}
\end{figure}

The actual forecasts are shown in Figure~\ref{fig:compare_z}.

\section{Conclusions}\label{sec:ConclusionsBVE}

This paper was devoted to the recreation of the first successful numerical weather forecast using machine learning. We have shown that the method based on physics-informed neural networks is capable of outperforming the original ENIAC integrations in terms of the mean error and RMSE, and matching it in terms of the S1 score. The method also outperforms the persistence prediction in terms of the RMSE and S1 score, while it could not consistently outperform the persistence prediction in terms of the mean error. 

We should like to stress that physics-informed neural networks have several advantages over classical numerical methods, which are particularly relevant for the atmospheric sciences. Firstly, the method avoids having to compute derivatives using numerical methods, since all derivatives are computed using automatic differentiation. This renders the method meshless which greatly simplifies solving differential equations on manifolds, and in particular on the sphere. This was a major complication for the original ENIAC integrations~\cite{char50a}, who first formulated the equations on the sphere but then solved it on a polar stereographic projection so that Cartesian coordinates could be used. In turn, no such complications arose in the present work. Moreover, meteorological equations such as the the barotropic vorticity equation, but also more complicated models such as the shallow-water equations using the vorticity--divergence formulation require the solution of a Poisson equation on the sphere, which is a rather costly endeavour if spherical harmonics have to be used. No Poisson equation has to be solved using our physics-informed neural network formulation, which greatly simplifies the numerical procedure involved.

While the solution of the barotropic vorticity equation for numerical weather forecasting is of mostly historical interest today, there are some interesting parallels between the early days of numerical weather forecasting and the state-of-the-art of deep learning-based solvers for meteorology. In particular, these issues concern the accuracy of the numerical method, the lack of conservation, and the time it takes them for completing a numerical integration. All of these issues were overcome in less than 50 years~\cite{baue15a} for traditional numerical solvers for the atmospheric governing equations. Given the immense interest in the field of scientific machine learning today and the trajectory of research in artificial intelligence in general, we are optimistic that it will not take 50 years for these issues to also be overcome for machine-learning based numerical solvers. With this in mind we would like to encourage the development of advanced PINNs for numerical weather forecasts.

\section*{Acknowledgements}

The authors thank Peter Lynch for making the \texttt{Matlab} code for the recreating of the ENIAC forecasts available online. This research was undertaken thanks to funding from the Canada Research Chairs program and the NSERC Discovery Grant program. The research of RB is funded by the Deutsche Forschungsgemeinschaft (DFG, German Research Foundation) -- Project-ID 274762653 -- TRR 181.

{\footnotesize\setlength{\itemsep}{0ex}

}

\end{document}